# Convolutional Neural Networks as a Model of the Visual System: Past, Present, and Future

Grace W. Lindsay, PhD
*Gatsby Computational Unit/Sainsbury Wellcome Centre*
*University College London*
gracewlindsay@gmail.com

## Abstract

Convolutional neural networks (CNNs) were inspired by early findings in the study of biological vision. They have since become successful tools in computer vision and state-of-the-art models of both neural activity and behavior on visual tasks. This review highlights what, in the context of CNNs, it means to be a good model in computational neuroscience and the various ways models can provide insight. Specifically, it covers the origins of CNNs and the methods by which we validate them as models of biological vision. It then goes on to elaborate on what we can learn about biological vision by understanding and experimenting on CNNs and discusses emerging opportunities for the use of CNNS in vision research beyond basic object recognition.

# Introduction

Computational models serve several purposes in neuroscience. They can validate intuitions about how a system works by providing a way to test those intuitions directly. They also offer a means to explore new hypotheses in an ideal experimental testing ground, wherein every detail can be controlled and measured. In addition models open the system in question up to a new realm of understanding through the use of mathematical analysis. In recent years, convolutional neural networks (CNNs) have performed all of these roles as a model of the visual system.

This review covers the origins of CNNs, the methods by which we validate them as models of the visual system, what we can find by experimenting on them, and emerging opportunities for their use in vision research. Importantly, this review is not intended to be a thorough overview of CNNs or extensively cover all uses of deep learning in the study of vision (other reviews may be of use to the reader for this [1, 2, 3, 4, 5, 75]). Rather, it is meant to demonstrate the strategies by which CNNs as a model can be used to gain insight and understanding about biological vision.

According to [6], "a functional model attempts only to match the outputs of a system given the same inputs provided to the system, whereas a mechanistic model attempts to also use components that parallel the actual physical components of the system". Using these definitions, this review is concerned with the use of CNNs as *mechanistic* models of the visual system. That is, it will be assumed and argued that, in addition to an overall match between outputs of the two systems, subparts of a CNN are intended to match subparts of the visual system.

## 1. Where convolutional neural networks came from

The history of convolutional neural networks threads through both neuroscience and artificial intelligence. Like artificial neural networks in general, they are an example of brain-inspired ideas coming to fruition through an interaction with computer science and engineering.

### 1.1 Origins of the model

In the mid-twentieth century, Hubel and Wiesel discovered two major cell types in the primary visual cortex (V1) of cats [7]. The first type, the simple cells, respond to bars of light or dark when placed at specific spatial locations. Each cell has an orientation of the bar at which it fires most, with its response falling off as the angle of the bar changes from this preferred orientation (creating an orientation 'tuning curve'). The second type, complex cells, have less strict response profiles; these cells still have preferred orientations but can respond just as strongly to a bar in several different nearby locations. Hubel and Wiesel concluded that these complex cells are likely receiving input from several simple cells, all with the same preferred orientation but with slightly different preferred locations (Figure 1, left).

In 1980, Fukushima transformed Hubel and Wiesel's findings into a functioning model of the visual system [8]. This model, the Neocognitron, is the precursor to modern convolutional neural networks. It contains two main cell types. The S-cells are named after simple cells and replicate their basic features: specifically, a 2-D grid of weights is applied at each location in the input image to create the S-cell responses. A "plane" of S-cells thus has a retinotopic layout with all cells sharing the same preferred visual features and multiple planes existing at a layer. The response of the C-cells (named after complex cells) is a nonlinear function of several S-cells coming from the same plane but at different locations.

After a layer of simple and complex cells representing the basic computations of V1, the Neocognitron simply repeats the process again. That is, the output of the first layer of complex cells serves as the input to the second simple cell layer, and so on. With several repeats, this creates a hierarchical model that mimics not just the operations of V1 but the ventral visual pathway as a whole. The network is "self-organized", meaning weights change with repeated exposure to unlabeled images.

By the 1990s, many similar hierarchical models of the visual system were being explored and related back to data [9]. One of the most prominent ones, HMAX, used the simple 'max' operation over the activity of a set of simple cells to determine the response of the C-cells and was very robust to image variations. Because these models could be applied to the same images used in human psychophysics experiments, the behavior of a model could be directly compared to the ability of humans to perform rapid visual categorization. Through this, a correspondence between these hierarchical models and the first 100-150ms of visual processing was found [123]. Such models were also fit to capture the responses of V4 neurons to complex shape stimuli [124].

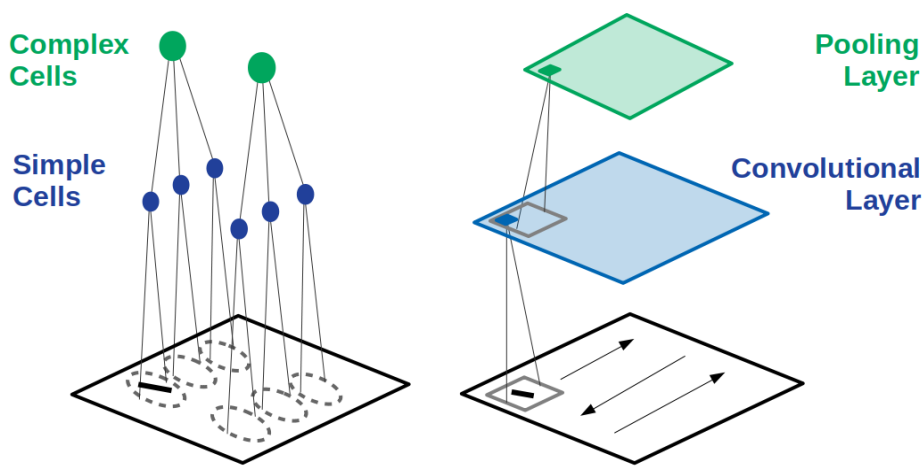

*Figure 1. The relationship between components of the visual system and the base operations of a convolutional neural network. Hubel and Wiesel discovered that simple cells (left, blue) have preferred locations in the image (dashed ovals) wherein they respond most strongly to bars of particular orientation. Complex cells (green) receive input from many simple cells and thus have more spatially invariant responses. These operations are replicated in a convolutional neural network (right). The first convolutional layer (blue) is produced by applying a convolution operation to the image. Specifically, the application of a small filter (gray box) to every location in the image creates a feature map. A convolutional layer has as many feature*

*maps as filters applied (only one shown here). Taking the maximal activation in a small section of each feature map (gray box) downsamples the image and leads to complex cell-like responses (green). This is known as a "max-pooling" operation and the downsampled feature maps produced by it comprise a pooling layer.*

**1.2 CNNs in computer vision**
The convolutional neural network as we know it today comes from the field of computer vision yet the inspiration from the work of of Hubel and Wiesel is clearly visible in it (Figure 1) [10]. Modern CNNs start by convolving a set of filters with an input image and rectifying the outputs, leading to "feature maps" akin to the planes of S-cells in the Neocognitron. Max-pooling is then applied, creating complex cell-like responses. After several iterations of this pattern, non-convolutional fully-connected layers are added and the last layer contains as many units as number of categories in the task, in order to output a category label for the image (bottom, Figure 2).

The first major demonstration of the power of CNNs came in 1989 when it was shown that a small CNN trained with supervision using the backpropagation algorithm could perform handwritten digit classification [11]. However, these networks did not really take off until 2012, when an 8-layer network (dubbed 'AlexNet', 'Standard Architecture' in Figure 3) trained with backpropagation far exceeded state-of-the-art performance on the ImageNet challenge. The ImageNet dataset is comprised of over a million real-world images, and the challenge requires classifying an image into one of a thousand object categories. The success of this network demonstrated that the basic features of the visual system found by neuroscientists were indeed capable of supporting vision; they simply needed appropriate learning algorithms and data.

In the years since this demonstration, many different CNN architectures have been explored, with the main parameters varied including network depth, placement of pooling layers, number of feature maps per layers, training procedures, and whether or not residual connections that skip layers exist [10]. The goal of exploring these parameters in the computer vision community is to create a model that performs better on standard image benchmarks, with secondary goals of making networks that are smaller or train with less data. Correspondence with biology is not a driving factor.

## 2. Validating CNNs as a model of the visual system
The architecture of a CNN has (by design) direct parallels to the architecture of the visual system. Images fed into these networks are usually first normalized and separated into three different color channels (RGB), which captures certain computations done by the retina. Each stacked bundle of convolution-nonlinearity-pooling can then be thought of as an approximation to a single visual area---usually the ones along the ventral stream such as V1, V2, V4 and IT---each with its own retinotopy and feature maps. This stacking creates receptive fields for individual neurons that increase in size deeper in the network and the features of the image that they respond to become more complex. As has been mentioned, when trained to, these architectures can take in an image and output a category label in agreement with human judgement.

All of these features make CNNs good candidates for models of the visual system. However, all of these features have been explicitly built into CNNs. To validate that CNNs are performing computations similar to those of the visual system, they should match the visual system in additional, non-engineered ways. That is, from the assumptions put into the model, further features of the data should fall out. Indeed, many further correspondences have been found at the neural and behavioral levels.

## 2.1 Comparison at the neural level

One of the major causes of the recent resurgence of interest in artificial neural networks amongst neuroscientists is the finding that they can recapitulate the representation of visual information along the ventral stream. In particular, when CNNs and animals are shown the same image (Figure 2), the activity of the artificial units can be used to predict the activity of real neurons, with accuracy beyond that of any previous methods.

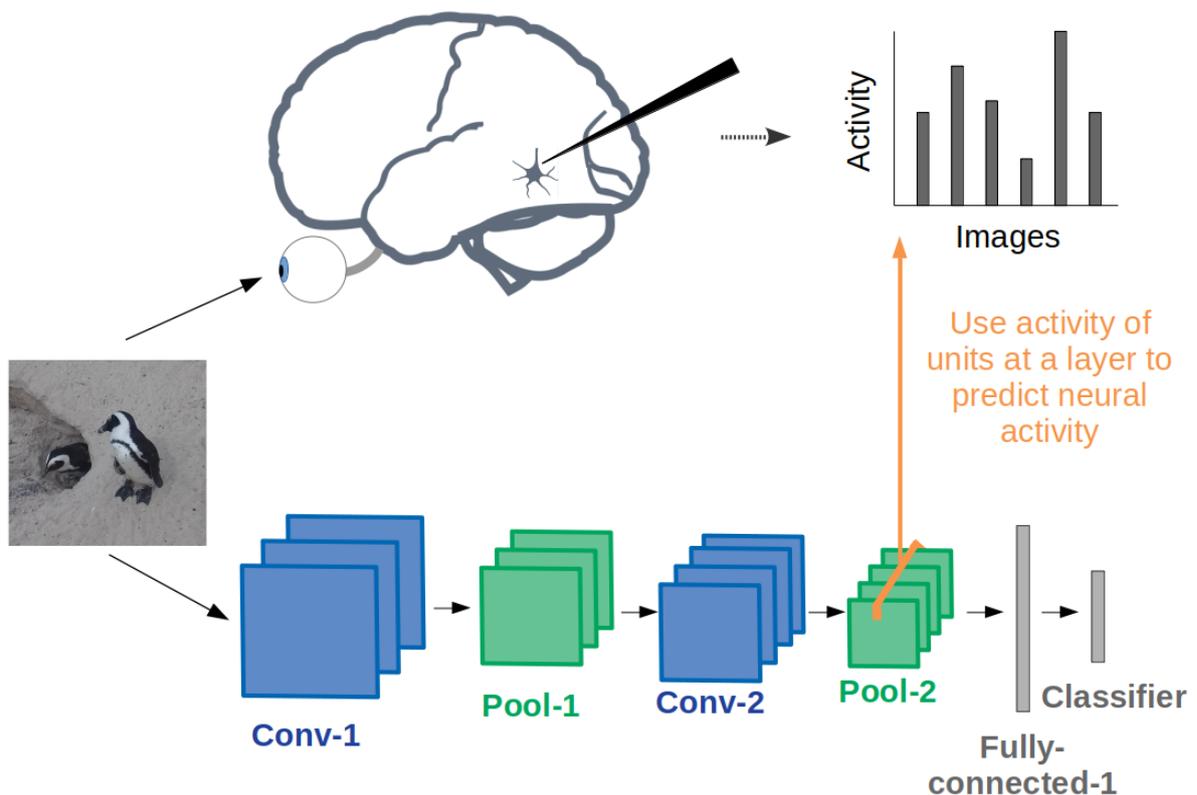

*Figure 2. Validating CNNs as a model by comparing their representations to those of the brain. A CNN trained to perform object recognition (bottom) contains several convolutional layers (Conv-1 and Conv-2 shown here, each followed by max-pooling layers Pool-1 and Pool-2) and ends with a fully-connected layer with a number of units equal to the number of categories. When the same image is shown to a CNN and an animal (top), the response from each can be compared. Here, the activity of single units recorded from a visual area in the brain can be predicted from the activity of artificial units at a particular layer.*

One of the early studies to show this [12], published in 2014, recorded extracellular activity in macaque during the viewing of complex object images. Regressing the activity of a real V4 or IT neuron onto the activity of the artificial units in a network (and cross-validating the predictive ability with a held-out test set), the authors found that networks that performed better on an object recognition task also better predicted neural activity (a relationship also found using video classification [13]). Furthermore, the activity of units from the last layer of the network best predicted IT activity and the penultimate layer best predicted V4. This relationship between models and the brain wherein later layers in the network better predict higher areas of the ventral stream has been found in several other studies, including using human fMRI [14], MEG [15], and with video instead of static images as the stimuli [16].

Another method to check for a correspondence between different populations is representational similarity analysis (RSA) [17]. This method starts by creating a matrix for each population that represents how dissimilar the responses of that population are for every pair of images. This serves as a signature of the population's representation properties. The similarity between two different populations is then measured as the correlation between their dissimilarity matrices. This method was used in 2014 [18] to demonstrate that the later layers of an AlexNet network trained on ImageNet match multiple higher areas of the human visual system, along with monkey IT, better than previously-used models.

Because dissimilarity matrices can be created from any kind of responses, including behavioral outputs, RSA is widely applicable as a means to compare data from different experimental methods and models. It is also a straightforward way to incorporate and compare full population responses, whereas regression techniques focus on a single neuron or voxel at a time. On the other hand, the regression techniques allow for selective weighting of the model features most relevant for fitting the data, which may be more informative than the more 'unsupervised' RSA approach. In all cases, details of the methodology and interpretation should be carefully considered [19, 20].

Many of the studies comparing CNNs to biological data have highlighted their ability to explain later visual areas such as V4 and IT. This is a notable feat of CNNs because the complex response properties of these areas have made them notoriously difficult to fit compared to primary visual cortex. Recent work, however, has shown that early-to-middle layers of task-trained CNNs can also predict V1 activity beyond the ability of more traditional V1 models [21].

Beyond predicting neural activity or matching overall representations, CNNs can also be compared to neural data using more specific features traditionally used in systems neuroscience. Similarities have been found between units in the network and individual neurons in terms of response sparseness and size tuning, but differences have been found for object selectivity and orientation tuning [22]. Other studies have explored tuning to shape and categories [23], and how responses change with changes in pose, location, and size [24, 25, 26].

Generally, the similarities with real neural activity that emerge from training a CNN to perform object recognition suggests that this architecture and task do indeed have some similarity to the architecture and purpose of the visual system.

**2.2 Comparison at the behavioral level**
Insofar as CNNs outperform any previous model of the visual system on real-world image classification, they can be considered a good match to human behavior. However, overall accuracy on the standard ImageNet task is only one measure of CNN behavior, and it is the one for which the network is explicitly optimized.

A deeper comparison can be made by looking at the errors these networks make. While there is only one way to be correct, there are many ways humans and models can make mistakes in classification, and this can provide rich insight into their respective workings. Confusion matrices, for example, are used to represent how frequently images from one category are classified as belonging to another and can be compared between models and animal behavior. A large-scale study showed that several different deep CNN architectures show similar matches to human and monkey object classification, though the models are not predictive down to the image level [27].

Other studies have explicitly asked subjects to rate similarity between images and compared human judgement to model representations [28,29]. In [30], for example, a large dataset was taken from the website "Totally Looks Like". While many similarity studies find good matches from CNNs, this study demonstrated more challenging elements of similarity that CNNs struggled to replicate.

In addition to similarity, other psychological concepts that have been tested in CNNs include typicality [31], Gestalt principles [32], and animacy [33]. In [127], a battery of tests inspired by findings in visual psychophysics were applied to CNNs and CNNs were found to be similar to biological vision according to roughly half of them.

Another way to probe animal and CNN behavior to is make the classification task more challenging by degrading image quality. Several studies have added various types of noise, occlusion, or blur to standard images and observed a decrease in classification performance [34,35,36,37,38]. Importantly this performance decrease is usually more severe in the CNNs than it is in humans, suggesting biological vision has mechanisms for overcoming degradation. These points of mismatch are thus important to identify in order to steer future research (for example, see section 3.1). A particular CNN architecture known as a capsule network [39] was shown to be more robust to degradation, however, it is not exactly clear how to relate the 'capsules' in this architecture to parts of the visual system.

Another emerging finding in the behavioral analysis of CNNs is their reliance on texture. While an argument has been made for CNNs as a model of human shape sensitivity [40], other studies have demonstrated that CNNs rely too much on texture and not enough on shape when classifying images [41, 42].

Interestingly, it is possible for certain deep networks to outperform humans on some tasks [43]. This highlights an important tension between the goals of computer vision and those of neuroscience: in the former exceeding human performance is desirable, in the latter it is still counted as a mismatch between the model and the data.

Something to keep in mind when studying CNN behavior is that standard feedforward CNN architectures are believed to represent the very initial stages of visual processing, before various kinds of recurrent processing can take place. Therefore, when comparing the behavior of CNNs to animal behavior, fast stimulus presentation and backward masking are advised, as these are known to prevent many stages of recurrent processing [44].

## 2.2 Other forms of validation

While neural and behavioral comparisons are the main methods by which CNNs are validated as models of the visual system, other approaches can provide further support.

Methods for visualizing the image features that drive units at different layers in the network (Figure 4) [45, 46] have revealed preferred visual patterns that align with those found in neuroscience. For example, the first layer in a CNN has filters that look like Gabors, while later layers respond to partial object features and eventually fuller features such as faces. This supports the idea that the processing steps in CNNs match those in the ventral stream.

CNNs have also been used to produce optimal stimuli for real neurons. Starting from the above-mentioned procedure for predicting neural activity with CNNs, stimuli that were intended to maximally drive a neuron's firing rate were produced [47]. The fact that the resulting stimuli were indeed effective at driving the neuron beyond its normal rates despite being unnatural and unrelated to the images the network was trained on further supports the notion that CNNs are capturing something fundamental about the visual processing stream.

In a related vein, studies have also used CNNs to decode neural activity in order to recreate the stimuli that are being presented to subjects [48].

## 3. What we learn from varying the model

The above-mentioned studies have focused mainly on standard feedforward CNN architectures, trained via supervised learning to perform object recognition. Yet, with full control over these models it is possible to explore many variations in datasets, architectures, and training procedures. By observing how these changes make the model a better or worse fit to data, we can gain insight into how and why certain features of biological visual processing exist.

### 3.1 Alternative datasets

The ImageNet dataset has proven very useful for learning a set of basic visual features that can be adapted to many different tasks in a way that previous smaller and simpler datasets such as MNIST and CIFAR-10 were not. However, it contains images where objects are the focus and as a result is best suited for studying object recognition pathways. To study scene processing areas such as the occipital place area (OPA), several studies have instead trained on scene images. In [49], the responses of OPA could be predicted using a network trained to classify scenes. What's more, the authors were able to relate the features captured by the network to navigational affordances in the scene. In [50] scene-trained networks were able to capture representations of scene size collected using MEG. Such studies can in general help to identify the evolutionarily- or developmentally-determined role of different brain areas based on the extent to which their representations align with networks trained on different specialized datasets. For an open dataset of fMRI responses to different image sets, see [51].

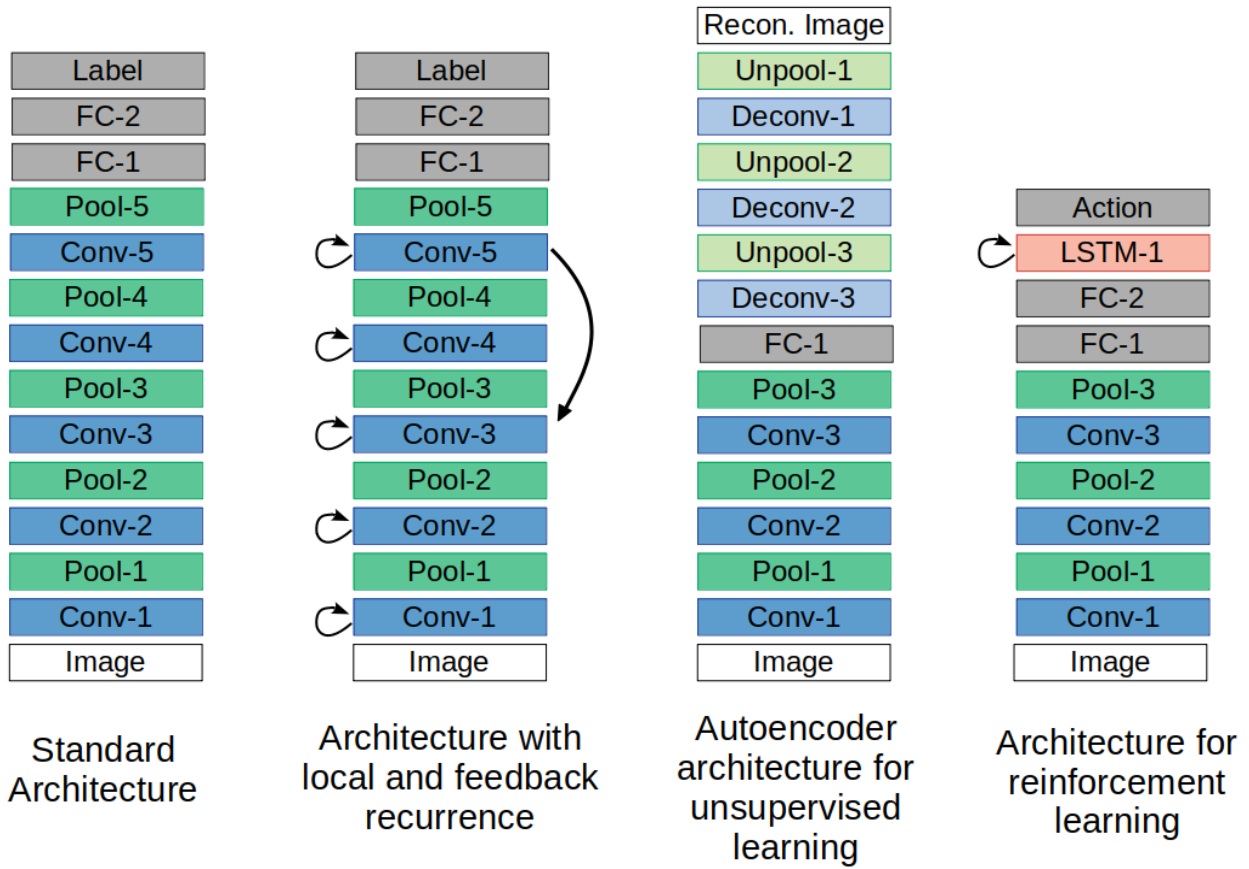

*Figure 3. Example architectures. The standard architecture shown here is similar to the original AlexNet, where an RGB image is taken as input and the activity at the final layer determines the category label. In the second architecture local recurrence is shown at each convolutional layer and feedback recurrence goes from layer 5 to layer 3. This network would be trained with the backpropoagation-through-time algorithm. The third network is an autoencoder. Here, the goal of the network is to recreate the input image. Crucially, the fully-connected layer is of significantly lower dimensionality than the image, forcing the network to find a more compact representation of relevant image statistics. Finally, an architecture for reinforcement learning is shown. Here, the aim of the network is to turn an input image representing information about the current state into an action for the network to take. This architecure is augmented with a specific type of local recurrent units known as LSTMs, which embue it with memory.*

Datasets have also been developed with the explicit intent of correcting ways in which CNNs do not match human behavior. For example, training with different image degradations can make networks more robust to those degradations [37]. However this only works for the noise model specifically trained and does not generalize to new noise types. In addition, in [42], a dataset that keeps abstract shapes but varies low-level textural elements was shown to decrease a CNN's texture bias. Whether animals have good visual performance due to an exposure to similarly diverse data during developmental or due to built-in priors instead remains to be determined.

**3.2 Alternative architectures**
A variety of purely feedforward architectures have been explored both in the machine learning community and by neuroscientists. In comparisons to data, AlexNet has been commonly used and performs well. But it can be beat by somewhat deeper architectures such as VGG models and ResNets [52,53]. With very deep networks---though they may perform better on image tasks---the relationship between layers in the network and areas in the brain may break down, making them a worse fit to the data. However in some cases the processing in very deep networks can be thought of as equivalent to the multiple stages of recurrent processing in the brain [54].

In primate vision, the first two processing stages---retina and the lateral geniculate nucleus---contain cells that respond preferentially to on-center or off-center light patterns, and only in V1 does orientation tuning become dominant. Yet, Gabor filters are frequently learned at the first layer of a CNN, creating a V1-like representation. Using a modified architecture wherein the connections from early layers are constrained as they are biologically by the optic nerve creates a CNN with on- and off-center responses at early stages and orientation tuning only later [55]. Thus the pattern of selectivity seen in primate vision is potentially a consequence of anatomical constraints.

In [56], various architectures were trained to perform a pair of auditory tasks (speech and music recognition). Through this the authors found that the two tasks could share three layers of processing before the network needed to split into specialized streams in order to perform well on both tasks. Recently, a similar procedure was applied to visual tasks and used to explain why specialized pathways for face processing arise in the visual system [57]. A similar question was also approached by training a single network to perform two tasks and looking for emergent sub-networks in the trained network [58]. Such studies can explain the existence of dorsal and ventral streams and other details of visual architectures.

Taking further inspiration from biology, many studies have explored the beneficial role of both local and feedback recurrence (Figure 3). Local recurrence refers to horizontal connections within a single visual area. By adding these connections to CNNs, studies have found that these recurrent connections make networks better at more challenging tasks [59, 60, 131, 132]. These connections can also help make the CNN representations a better match to neural data, particularly for challenging images and at later time points in the response [54, 61, 62, 63, 64]. These studies make a strong argument for the computational role of these local connections.

Feedback connections go from frontal or parietal areas to regions in the visual system or from higher visual areas back to lower areas [133]. Like horizontal recurrence, they are known to be common in biological vision. Connections from frontal and parietal areas are believed to implement goal-directed selective attention. Such feedback has been added to network models to implement cued detection tasks [103,134]. Feedback from higher visual areas back to lower ones are thought to implement more immediate and general image processing such as denoising. Some studies have added these connections in addition to local recurrence, and found that they can aid performance as well [65,66]. A study comparing different feedback and feedforward architectures suggests that feedback can help in part by increasing the effective receptive field size of cells [135].

### 3.3 Alternative training procedures

Supervised learning using backpropagation is the most common method of training CNNs, however other methods have the potential to result in a good model of the visual system. The 2014 study that initially showed a correlation between performance on object recognition and ability to capture neural responses [12], for example, did not use backpropagation but rather a modular optimization procedure.

Unsupervised learning, wherein networks aim to capture relevant statistics of the input data rather than match inputs to outputs, can also be used to train neural networks [67] (Figure 3). These methods may help identify a low-dimensional set of features that underlie high-dimensional visual inputs, and thus allow an animal to make better sense of the world and possibly build useful causal models [130]. Furthermore, due to the large amount of labeled data points required for supervised training, it is assumed that the brain must make use of unsupervised learning. However, as of yet, unsupervised methods do not produce models that capture neural representations as well as supervised methods. Behaviorally, a generative model was shown to perform much worse than supervised models on capturing human image categorization [128].. In addition, a model trained based on the concept of predictive coding [68] was able to predict object movement and replicate motion illusions [69]. Overall, the limits and benefits of unsupervised training for vision models needs further exploration.

Interestingly, a recent study found a class-conditioned generative model to be in better agreement with human judgment on controversial stimuli than models trained to perform classification [129]. While this generative model still relied on class labels for training and was thus not unsupervised, its ability to replicate aspects of human perception supports the notion that the visual system aims in part to capture a distribution of the visual world rather than merely funnel images into object categories.

A compromise between unsupervised and supervised methods is 'semi-supervised' learning, which has recently been explored as a means of making more biologically realistic networks [70, 71].

Reinforcement learning is the third major class of training in machine learning. In these systems, an artificial agent must learn to produce action outcomes in response to information from the environment, including rewards. Several such artificial systems have used convolutional architectures on the front end in order to process visual information about the world (Figure 3) [72,73,74]. It would be interesting to compare the representations learned in the context of these models to those trained by other mechanisms, as well as to data.

A simple way to understand the importance of training for a network is to compare to a network with the same architecture but random weights. In [32], the ability of a network to perform perceptual closing existed only in networks that had been trained with natural images, not random networks.

The above methods rely on training data, such as a set of images, that does not come from neural recordings. However, it is possible to train these architectures to replicate neural activity directly. Doing so can help identify the features of the input image most responsible for a neuron's firing [21, 76, 77]. Ideally, the components of the model can be also related back to anatomical features of the circuit in question [78,79] as is done when networks are trained on classification tasks.

A final hybrid option is to train on a classification task while also using neural data to constrain the intermediate representations to be brain-like [80], resulting in a network that better matches neural data and can perform the task.

## 4. How to understand CNNs

Varying a network's structure and training is one way to explore how it functions. Additionally, trained networks can be probed directly using techniques normally applied to brains, or those only available in models [81]. In either case, if we believe CNNs have been validated as a model of the visual system, any insight gained from probing how they work may apply to biological vision as well.

### 4.1 Empirical methods

The tools of the standard neuroscientific toolbox---lesions, recordings, anatomical tracings, stimulations, silencing, etc.---are all readily available in artificial neural networks [82] and can be used to answer questions about the workings of these networks.

"Ablating" individual units in a CNN, for example, can have an impact on classification accuracy, however, the impact of ablating a particular unit does not have a strong relationship with the unit's selectivity properties [83, 84].

In [49] images were manipulated or occluded to determine which features were responsible for the CNN's response to scenes, akin to how the function of real neurons is explored.

Other studies [60, 55] have analyzed the connectivity properties of trained networks to see if they replicate features of visual cortical anatomy.

One conceptual framework to describe what the stages of visual processing are doing is that of "untangling". High level concepts that are intertangled in the pixel or retinal representation get pulled apart to form easily separable clusters in later representations. This theory has been developed using biological data [85], however, recently, techniques for describing the geometry of these clusters (or manifolds) have been developed and used to understand the untangling process in deep neural networks [86, 87]. This work highlights the relevant features of these manifolds for classification and how they change through learning and processing stages. This can help identify which response features to look for in data.

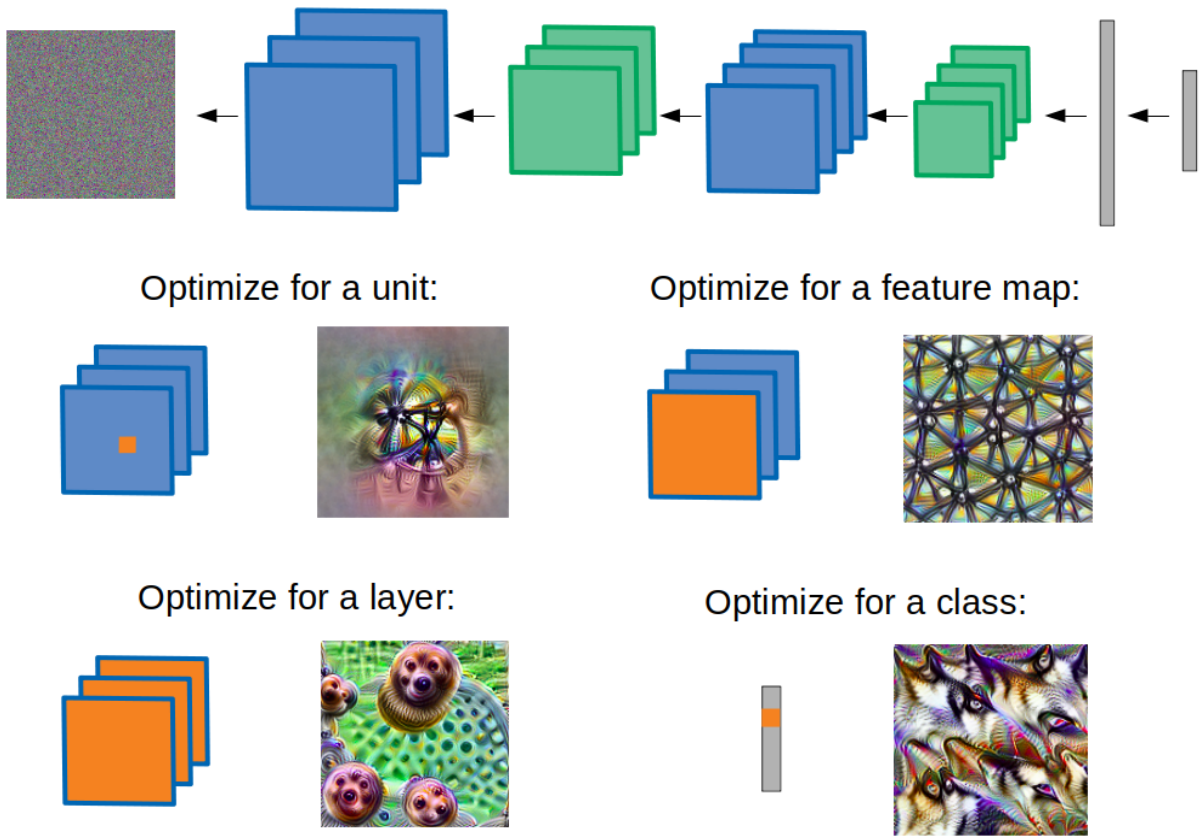

*Figure 4. Visualizing preferences of different network components. The general method for creating these images is shown at the top. Specifically, a component of the model is selected for optimization and gradient calculations sent back through the network indicate how pixels in an initially-noisy image should change in order to best activate the chosen component. Components can be individual units, whole feature maps, layers (using the DeepDream algorithm), or units representing a specific class in the final layer. Feature visualizations taken from [46].*

## 4.2 Mathematical Analyses

Given our full access to the equations that define a CNN, many mathematical techniques not currently applicable to brains can be performed. One common such tool is the calculation of gradients. Gradients indicate how certain components in the network affect others, potentially far away. They are used to train these networks by determining how a weight at one layer can be changed to decrease error at the output. However they can also be used to visualize the preferred features of a unit in a network. In that case, the gradient calculation goes all the way through to the input image, where it can be determined how individual pixels should change in order to increase the activity of a particular unit [46] (Figure 4). Multiple variants of feature visualization have also been used to probe neural network functions [89], for example, by showing which invariances exist at different layers [90].

Gradients can also be used to determine the role an artificial neuron plays in a classification task. In [88], it was found that the role a unit plays in classifying an image as being of a certain category is not tightly correlated with how strongly it responds to images of that category. Much like the ablation study above, this demonstrates a disassociation between tuning and function that should give neuroscientists pause about how informative a tuning-based analysis of the brain is.

Machine learning researchers and mathematicians also aim to cast CNNs in a light that makes them more amenable to traditional mathematical analysis. The concepts of information theory [91] and wavelet scattering, for example, have been used towards this end [92]. Another fruitful approach to mathematical analysis has been to study deep linear neural networks as this approximation makes more analyses possible [93].

Several studies in computational neuroscience have trained simple recurrent neural networks to perform tasks, and then interrogated the properties of the trained networks for clues as to how they work [94, 95, 96]. Going forward, more sophisticated techniques for understanding the learned representations and connectivity patterns of convolutional neural networks should be developed. This can both provide insight into how these networks work as well as indicate which experimental techniques and data analysis methods would be fruitful to pursue.

**4.3 Are they understandable?**
For some researchers, CNNs represent the unavoidable tradeoff between complexity and interpretability. To have a model complex enough to perform real-world tasks, we must sacrifice the desire to make simple statements about how each stage of it works---a goal inherent in much of systems neuroscience. For this reason, an alternative way to describe the network that is compact without relying on simple statements about computations has been proposed [97,126]; this viewpoint focuses on describing the architecture, optimization function, and learning algorithm of the network---instead of attempting to describe specific computations---because specific computations and representations can be seen as simply emerging from these three factors.

It is true that the historical aim of language-based descriptions of the roles of individual neurons or groups of neurons (e.g., 'tuned to orientation', or 'face detectors') seems woefully incomplete as a way to capture the essential computations of CNNs. Yet it also seems that there are still more compact ways to describe the functioning of these networks and that finding these simpler descriptions could provide a further sense of understanding. Only certain sets of weights, for example, allow a network to perform well on real-world classification tasks. As of yet, the main thing we know about these 'good' weights is that they can be found through optimization procedures. But are there essential properties we could identify that would result in a more condensed description of the network? The 'lottery ticket' method of training can produce networks that work as well as dense networks using only a fraction of the weights and a recent study noted that as many as 95% of model weights could be guessed from the remaining 5% [125,98]. Such findings suggest that more condensed (and thus potentially more understandable) descriptions of the computations of high-performing networks are possible.

A similar analysis of architectures could be done to determine which broad features of connectivity are sufficient to create good performance. Recent work using random weights, for example, helps to isolate the role of architecture versus specific weight values [99]. A more compact description of the essential features of a high-performing network is a goal for both machine learning and neuroscience.

One undeniably non-compact component of deep learning is the the dataset. As has been mentioned, large, real-world datasets are required to match the performance and neural representations of biological vision. Are we doomed to carry around the full ImageNet dataset as part of our description of how vision works? Or can a set of sufficient statistics be defined? Natural scene statistics have been a historically large part of both computational neuroscience and computer vision [100]. While much of that work has focused on lower order correlations that are insufficient to capture the relevant features for object recognition, the time seems ripe for explorations of higher order image statistics, particularly as advances in generative modeling [101] point to the ability to condense full and complex features of natural images into a model.

In any case, nearly all the critiques against the interpretability of CNNs could equally apply to biological networks. Therefore, CNNs make a good testing ground for deciding what understanding looks like in neural systems.

## 5. Beyond the Basics

Since 2014, an explosion of research has answered many questions about how varying different features of a CNN can change its properties and its ability to match data. These findings, in turn, have aided progress in understanding both the 'how' and 'why' of biological vision. Beyond this, having access to an 'image computable' model of the visual system opens the door to many more modeling opportunities that can explore more than just core vision.

### 5.1 Exploring cognitive tasks

The relationship between image encoding and memorability was explored in [102]. The authors showed that the overall magnitude of the response of later layers of a CNN correlated with how memorable images were found to be experimentally.

In [102], an attractor network based on semantic features was added to the end of a CNN

architecture. This additional processing stage was able to account for perirhinal cortical activity during a semantic task.

Visual attention is known to enhance performance on challenging visual tasks. Applying the neuro-modulatory effects of attention to the units in a CNN was shown to increase performance in these networks as well, more so when applied at later rather than earlier layers [88]. This use of task-performing models has also led to better theories of how attention can work than those stemming solely from neural data [103].

Finally, CNNs were also able to recapitulate several behavioral and neural effects of fine-grain perceptual learning [104].

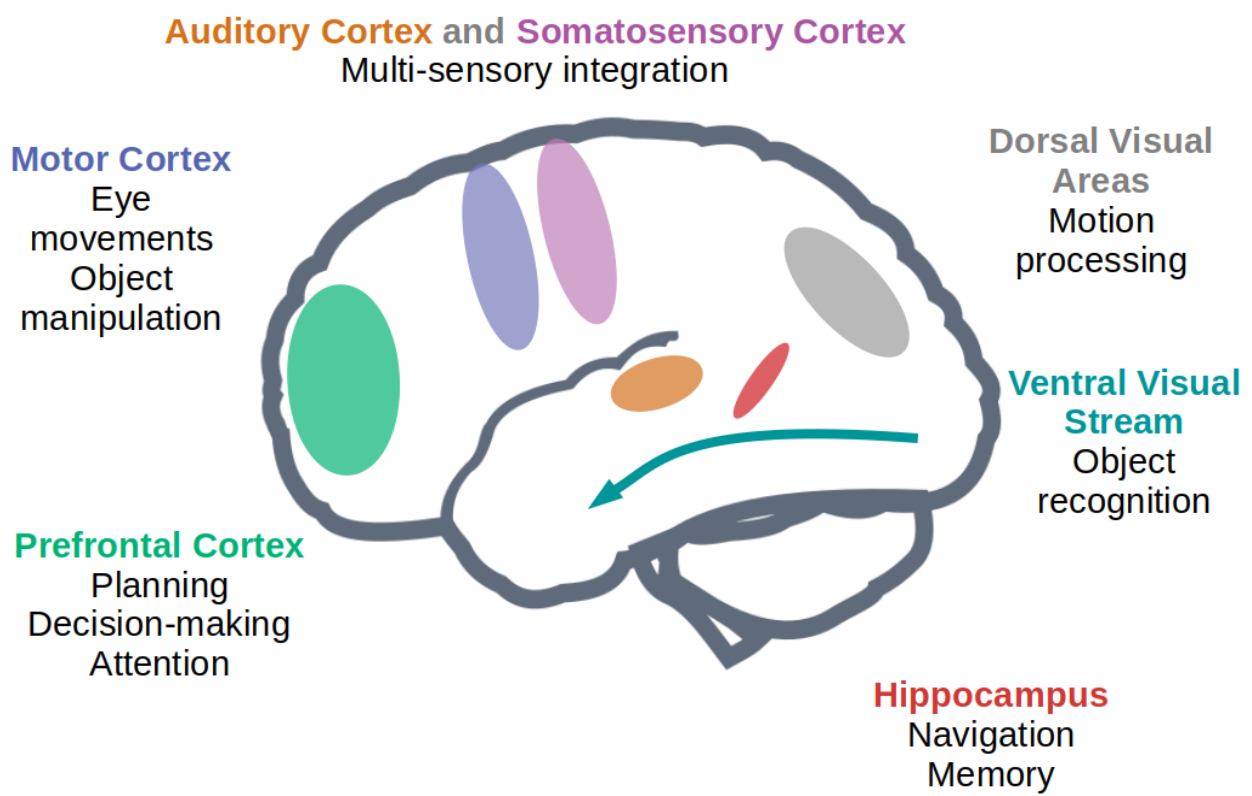

*Figure 5. A sampling of the many uses of visual information by the brain. Much of the use of CNNs as a model of visual processing has focused on the ventral visual stream. However, the visual system interacts with many other parts of the brain to achieve many different goals. Future models should work to integrate CNNs into this larger picture.*

### 5.2 Adding biological details
Some of the architectural variants described above---particularly the addition of local and feedback recurrence---are biologically-inspired and assumed to enhance the ability of the network to perform challenging tasks. Other brain-inspired details have been explored by the machine learning community in the hopes that these additions would be useful for difficult tasks. This includes foveation and saccading [105] for image classification.

Another reason to add biological detail would be out of a belief that it may make the network *worse*. The fact that CNNs lack many biological details does not make them a poor model of the visual system necessarily; it simply makes them an abstract one. However, it should be an ultimate aim to bring abstract and detailed models together to show how the high-level computations of a CNN are implemented using the machinery available to the brain. To this end, work has been done on spiking CNNs [106] and CNNs with stochastic noise added [61]. These attempts can also identify aspects of these biological details that are useful for computation.

The long history of modeling the circuitry of visual cortex can provide more ideas about what details to incorporate and how. A pre-existing circuit model of V1 anatomy and function [107], for example, was placed into the architecture of a CNN and used to replicate effects of visual attention [108]. A more extreme approach to adding biological detail can be found in [109], where the connectome of the fly visual system defined the architecture of the model, which was then trained to perform visual tasks.

## 5. Limitations and future directions

As with any model, the current limitations and flaws of CNNs should point the way to future research directions that will bring these models more in line with biology.

The basic structure of a CNN assumes weight sharing. That is, a feature map is the result of the exact same filter weights applied at each location in the layer below. While selectivity to visual features like orientation can appear all over the retinotopic map, it is clear that this isn't the result of any sort of explicit weight sharing. Either genetic programming ensures the same features are detected throughout space, or this property is learned through exposure. Studies on "translation tolerance" have shown the latter may be true [110]. Weight sharing makes CNNs easier to train, however ideally the same results could be found using a more biologically-plausible way of fitting filters.

Furthermore, in most CNNs Dale's law is not respected. That is, the same neuron can be weighted by both inhibitory (negative) and excitatory (positive) weights. In the visual system, connections between areas tend to come only from excitatory cells. To be consistent with biology, a negative feedforward weight could be interpreted as an excitatory feedforward connection that acts on local inhibitory neurons. But this relationship between excitatory feedforward connections and the need for local inhibitory recurrence points to a complication of adding biological details to these networks: some of these biological details may only function well or make sense in light of others and thus need to be added together.

To some, the way in which these networks are trained also poses a problem. The backpropagation algorithm is not considered biologically-plausible enough to be an approximation of how the visual system actually learns. However most methods for model fitting in computational neuroscience do not intend to mimic biological learning, and backpropagation could be thought of as just another parameter fitting technique. That being said, several researchers are investigating means by which the brain could perform something like backpropagation[111, 112, 113, 114]. Comparing models trained using more biologically plausible techniques to standard supervised learning (as well as to the unsupervised and reinforcement learning approaches discussed above) could offer insights as to the role of learning in determining representations.

The vast majority of studies comparing CNNs to biological vision have used data from humans or non-human primates. Where attempts have been made to compare CNNs to one of the most commonly used animal groups in neuroscience research---rodents---results are not nearly as strong as they are for primates [115, 116, 136]. Understanding what can turn CNNs into a good model of rodent vision would go a long way in understanding the difference between primate and rodent vision and would open rodent vision up to the exploration tactics described here. CNNs have also been compared to the behavioral patterns of pigeons on a classification task [117].

Even in the context of primate vision, simple object or scene classification tasks only represent a small fraction of what visual systems are capable of and used for naturally. More ethologically-relevant and embodied tasks such as navigation, object manipulation and visual reasoning, may be needed to capture the full diversity of visual processing and its relation to other brain areas. Early versions of this idea are already being explored [118,119 ]. The study of insect vision has historically taken this more holistic approach and may make for useful inspiration [120].

## 6. Conclusions

The story of convolutional neural networks started with a study on the tuning properties of individual neurons in primary visual cortex. Yet one of the impacts of using CNNs to study the visual system has been to push the field away from focusing on interpretable responses of single neurons and toward population-level descriptions of how visual information is represented and transformed in order to perform visual tasks. The shift towards models that actually *do* something has forced a reshaping of the questions around the study of the visual system. Neuroscientists are adapting to this new style of explanation and the different expectations that come with it [121].

Importantly, these models have also made it possible to reach some of the pre-existing goals in the study of vision. In 2007 [122], for example, a perspective piece on the study of object recognition claimed that "Progress in understanding the brain's solution to object recognition requires the construction of artificial recognition systems that ultimately aim to emulate our own visual abilities, often with biological inspiration" and that "instantiation of a working recognition system represents a particularly effective measure of success in understanding object recognition." In this way, CNNs as a model of the visual system are a success.

Of course, nothing can be learned about biological vision through CNNs in isolation, but rather only through iteration. The insights gained from experimenting with CNNs should shape future experiments in the lab, which in turn, should inform the next generation of models.

## Acknowledgements

Thanks to SciDraw.io for providing brain and neuron drawings. Feature visualizations in Figure 4 were taken from [46] are licensed under Creative Commons Attribution CC-BY 4.0. This work was supported by a Marie Skłodowska-Curie Individual Fellowship and a Sainsbury Wellcome Centre/Gatsby Computational Unit Research Fellowship.